\documentclass[aps,pre,twocolumn,showpacs,preprintnumbers,floatfix]{revtex4-2}
\usepackage{amsmath,amssymb,amsfonts}
\usepackage{graphicx}
\usepackage{xargs}          
\usepackage{hyperref}

\newcommand{\rp}{\rho/\mu}

\begin{document}

\title{Estimating Sloppy Directions via KDE: The Case of Kirman's Ants}

\author{Karl Naumann-Woleske}
\email{karl.naumann-woleske@wu.ac.at}
\affiliation{Vienna University of Economics and Business (WU), Vienna, Austria\\Econophysics Lab, Institut Louis Bachelier, 28 Pl. de la Bourse, Palais Brongniart, 75002 Paris, France}

\date{\today}

\begin{abstract}
Models whose predictions depend on only a handful of well-constrained parameter combinations, termed sloppy models, are ubiquitous in nonlinear stochastic systems.
The information-geometric approach to sloppiness advocates using the symmetrized Kullback--Leibler divergence and its associated Hessian, the Fisher Information Matrix (FIM), as the natural loss function.
However, prior applications have relied on analytically known or parametrically fitted distributions.
In practice, for general agent-based or stochastic models the distribution must be estimated from simulation data.
I demonstrate, using Kirman's ant recruitment model as a worked example, that a standard kernel density estimate (KDE) converges to the analytical FIM eigenvectors and eigenvalues with simulation budgets accessible in practice.
I derive the analytical Hessian in closed form, show numerical convergence of the KDE-based estimate as a function of simulation data, and extend the analysis to fixed-lag joint densities, where the FIM is full rank and the second eigenvalue is small but non-zero.
\end{abstract}

\maketitle

\section{Introduction}

Confronted with highly parameterized models and a limited set of observations, scientists can nonetheless make accurate predictions.
The reason is that many models are \emph{sloppy}: their predictions depend only on a handful of \emph{stiff} parameter combinations, while the remaining \emph{sloppy} combinations have little to no effect on the predicted outcomes~\cite{BrownSethna2003StatisticalMechanicalApproaches,BrownEtAl2004StatisticalMechanicsComplex}.
The notion of stiff and sloppy parameter directions was first introduced by Brown and Sethna~\cite{BrownSethna2003StatisticalMechanicalApproaches} while studying models of biochemical regulation.
For each of the individual parameter estimates, ``95\% confidence intervals each spanned more than a factor of 50''~\cite{GutenkunstEtAl2007UniversallySloppyParameter}, yet the models retained accurate predictive power because the stiff directions were well-constrained by data.
Sloppiness has since been identified across cell signaling, radioactive decay, neural networks, quantum wave functions, the Ising model, and macroeconomic agent-based models~\cite{QuinnEtAl2023InformationGeometryMultiparameter,WaterfallEtAl2006SloppyModelUniversalityClass,MachtaEtAl2013ParameterSpaceCompression}.

The standard framework identifies stiff and sloppy directions via the eigendecomposition of the Hessian matrix of a loss function that measures the change in model predictions when parameters are perturbed.
Two loss functions dominate the literature: the mean-squared error (MSE) on predicted time series, and the symmetrized Kullback--Leibler (sKL) divergence between predicted distributions~\cite{KullbackLeibler1951InformationSufficiency,Jeffreys1948TheoryProbability}.
The Hessian of the sKL divergence equals the FIM~\cite{QuinnEtAl2023InformationGeometryMultiparameter}, a natural Riemannian metric on the manifold of model predictions~\cite{TranstrumEtAl2011GeometryNonlinearLeast,TranstrumEtAl2010WhyAreNonlinear}.
Comparing distributions rather than time series also reduces the sensitivity to stochastic path noise and makes the distributional approach preferable for nonlinear stochastic models.

Yet existing worked examples have a common feature: the distribution being compared is either known analytically~\cite{MachtaEtAl2013ParameterSpaceCompression} or fit to a known parametric form~\cite{GutenkunstEtAl2007UniversallySloppyParameter}.
In practice, for complex agent-based models (ABMs) the stationary distribution is unknown and must be estimated nonparametrically from simulation data.
Prior work on sloppiness in ABMs ~\cite{Naumann-WoleskeEtAl2025ExplorationParameterSpace} applied the sloppy-models framework to a high-dimensional macroeconomic ABM but relied on the MSE loss for stochastic time-series observables rather than a distributional loss.
This leads to high uncertainty in the estimated stiff and sloppy directions, which in turns requires a large simulation budget, suggesting that an alternative distributional approach may be preferable but in this case there is no analytical prior on the distribution of outcomes.
To my knowledge, whether a nonparametric estimate of the distribution recovers the true set of stiff and sloppy directions, i.e. the eigendecomposition of the FIM, has not been demonstrated.

I address this gap using Kirman's ant recruitment model~\cite{Kirman1993AntsRationalityRecruitment}, a simple non-trivial stochastic model with a known analytical stationary distribution.
With only two parameters $(\rho, \mu)$, the model has a closed-form FIM, providing an exact ground truth.
I show that a Gaussian KDE with a logit transform, combined with central finite differences in log-parameter space, recovers the analytical eigenvectors at modest simulation budgets, and the eigenvalues once the kernel bandwidth is chosen appropriately.
In this case the sloppy direction is an exact symmetry, since scaling $\rho$ and $\mu$ together reparameterises time and leaves the stationary distribution unchanged, giving an analytical eigenvalue of exactly zero rather than the small non-zero eigenvalues that terminate a classic sloppy spectrum.
The numerical estimation pipeline is unaware of this degeneracy, resulting in a non-zero second eigenvalue due to KDE smoothing and sampling noise. The pipeline recovers the degeneracy as the second eigenvalue shrinks as a function of the simulation budget.
To investigate a non-degenerate case with two non-zero eigenvalues, I repeat the experiment using fixed-lag joint densities, which have an exponential covariance, and thus a full-rank FIM. 
\citet{MoranEtAl2020SchrodingersAntsContinuous} provide the basis for writing the functional form of the joint distribution, from which I derive the structure of the joint-density FIM and its eigendecomposition.
Here too, I find that the numerical KDE-based approach to estimating the sKL-based FIM succeeds in recovering the analytically suggested directions, together with both the stiff and sloppy eigenvalues.

The ability of such a simple estimation pipeline to recover the geometry of the loss landscape can be used as a means of exploring the model's parameter space efficiently, by following the stiff direction to maximize the variation of dynamics observed along the lines of \citet{Naumann-WoleskeEtAl2025ExplorationParameterSpace}.

\section{Kirman's Ant Recruitment Model}

Consider a set of $N$ ants facing two equivalent food sources $A$ and $B$, where $k$ is the number of ants at source $A$.
Each period, a randomly drawn ant may switch its food source with probability $\rho \in [0,1]$ (random switching), or alternatively recruit another randomly drawn ant with probability $\mu \in [0,1]$ (herding).
Let $x_t = k/N$ denote the fraction of ants at source $A$ at time $t$.

\begin{figure}[htpb]
    \centering
    \includegraphics[width=\columnwidth]{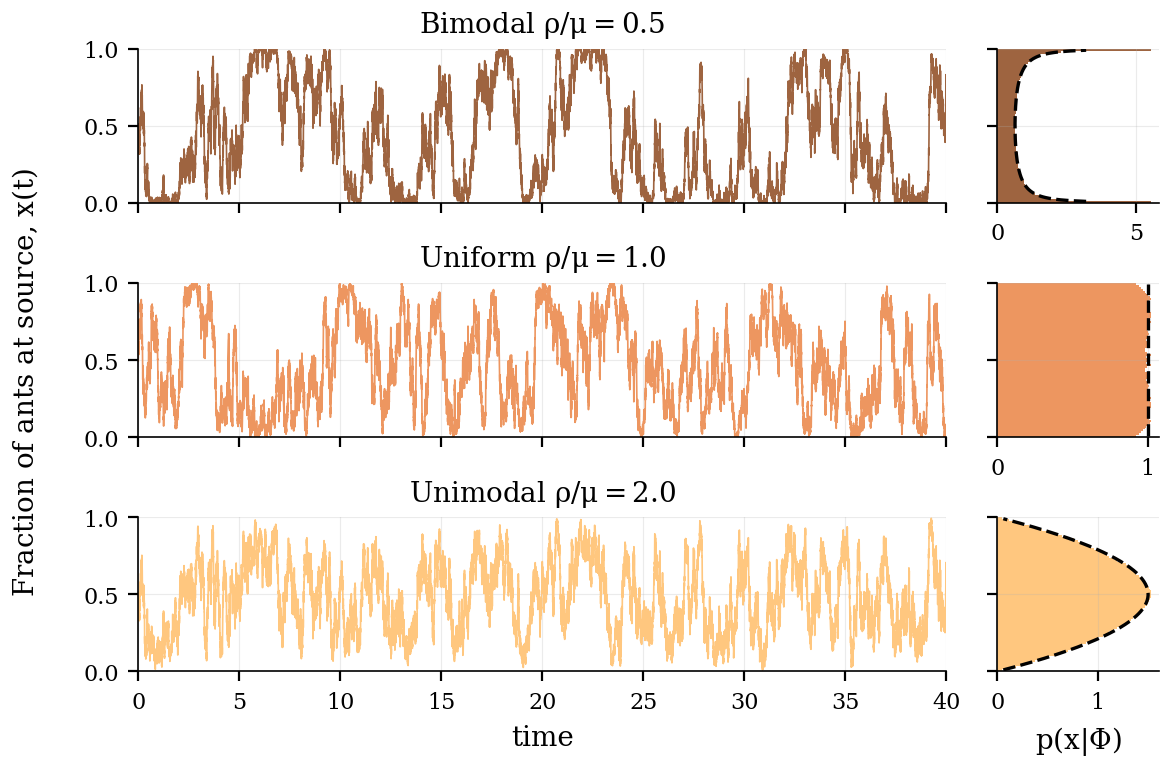}
    \caption{Dynamics of Kirman's ant model via
        Eq.~\eqref{eq:ants_discrete_dynamics} with $\Delta t=10^{-4}$, run for $T=10^{3}$ time units ($10^{7}$ steps) and $x_0=\tfrac{1}{2}$.
        Left: time series for 40 steps (including all $\Delta t$ substeps).
        Right: stationary density $p(x|\Phi)$ of the ants. Dashed lines represent the analytical $\text{Beta}(\rp,\rp)$ densities for the respective $\rp$ ratio.
        Parameters: $\mu=1.0$ fixed; $\rho=0.5$ (bimodal,
        $\rho/\mu<1$), $\rho=1.0$ (uniform, $\rho/\mu=1$),
        $\rho=2.0$ (unimodal, $\rho/\mu>1$).}
    \label{fig:ant_dynamics}
\end{figure}

Following~\citet{Kirman1993AntsRationalityRecruitment} and~\citet{MoranEtAl2020SchrodingersAntsContinuous}, in the $N\to \infty$ limit the dynamics of $x$ obey the It\^o stochastic differential equation
\begin{equation}\label{eq:ants_continuous_dynamics}
    dx_t = \rho(1-2x_t)\,dt + \sqrt{2\mu\, x_t(1-x_t)}\,dW_t,
\end{equation}
where $W_t$ is a standard Wiener process.
This discretizes via Euler--Maruyama with step $\Delta t$ as
\begin{align}\label{eq:ants_discrete_dynamics}
    x_{t} - x_{t-\Delta t} =&~ \rho(1-2x_{t-\Delta t})\,\Delta t\nonumber\\
        &+ \sqrt{2\mu\, x_{t-\Delta t}(1-x_{t-\Delta t})}\,\sqrt{\Delta t}\;\xi_t,
\end{align}
with $\xi_t \sim \mathcal{N}(0,1)$ i.i.d.
Figure~\ref{fig:ant_dynamics} shows representative dynamics and stationary distributions for the three regimes.

\subsection*{Stationary distribution}
For $\rho > 0$, the normalized stationary distribution is a symmetric Beta distribution~\cite{Kirman1993AntsRationalityRecruitment,MoranEtAl2020SchrodingersAntsContinuous}:
\begin{equation}\label{eq:ants_stationary_distribution}
    P(x|\Phi) = \frac{\Gamma(2\rp)}{\Gamma^2(\rp)}\,
        \bigl[x(1-x)\bigr]^{\rp-1}
        \;\sim\; \mathrm{Beta}\!\left(\rp,\,\rp\right).
\end{equation}
The ratio $\rp$ determines the qualitative behavior of the model: for $\rp < 1$ the distribution is bimodal (density diverging at the boundaries), for $\rp > 1$ it is unimodal (peaked at $x = \tfrac{1}{2}$), and $\rp = 1$ yields a uniform distribution.
The ratio $\rp$ therefore controls the qualitative shape of the stationary density, with the change from bimodal to unimodal occurring at $\rp = 1$.
In this case, any steps keeping $\rp$ constant correspond, trivially, to a reparameterization of time in the model, encoding a simple symmetry in the model.

\subsection*{Lag $\tau$ joint distribution}
The stationary distribution depends on $\rho$ and $\mu$ only through their ratio, so scaling both together leaves it unchanged.
To contrast this, I also consider the joint density of a pair $(x_t, x_{t+\tau})$, where the density not only depends on the ratio (via the stationary distribution) but also separately on the relaxation timescale that \citet{MoranEtAl2020SchrodingersAntsContinuous} show to depend only on the random switching probability $\rho$.

Following \citet[][Eq. 18]{MoranEtAl2020SchrodingersAntsContinuous}, the probability density of $x_{t+\tau}$ conditional on starting at $x_t$ with $\tau$ periods earlier is given by
\begin{align}
    P(x_{t+\tau}|x_t,\Phi) = P(x_{t+\tau}|\Phi) + \sum_{n\geq1}e^{-\mu\mathcal{E}_n\tau}\Psi_n(x_t)f_n(x_{t+\tau})\label{eq:joint_propagator},
\end{align}
where the first term is the stationary density of Eq. \eqref{eq:ants_stationary_distribution}, and each term of the sum is a relaxation mode of shape $f_n(x_{t+\tau})$ with amplitude $\Psi_n(x_t)$ given by the starting point $x_t$, and a relaxation rate $\mu\mathcal{E}_n=n(2\rho + \mu(n-1))$.

The joint density $P(x_t,x_{t+\tau})$ is consequently given by
\begin{align}
    P(x_t, x_{t+\tau}) =&~ P(x_{t})P(x_{t+\tau})\nonumber\\ 
                             &+ \sum_{n\geq1}e^{-\mu\mathcal{E}_n\tau}P(x_{t})\Psi_n(x_t)f_n(x_{t+\tau}),
\end{align}
depending on the stationary and convergence components.
Clearly, as $\tau\to\infty$, the points are independent and joint density relaxes to the product of the stationary distribution, thus depending only on $\rp$.
Convergence is exponential with the slowest mode being $n=1$, such that for any finite $\tau$ the density also depends on $\mu\mathcal{E}_1=2\rho$.

In the case of $n=1$, the joint density can be expressed analytically as 
\begin{align}
    P(x_t, x_{t+\tau}) =&~ P(x_{t})P(x_{t+\tau})\left[1 + e^{-2\rho \tau}\frac{(x_t-\frac12)(x_{t+\tau}-\frac12)}{V}\right],\label{eq:joint_density_n1}
\end{align}
where $V$ is the variance of the stationary distribution of Eq. \eqref{eq:ants_stationary_distribution} (see Appendix \ref{appx:joint_density_n1} for a derivation).

\section{Analytical Hessian of the sKL Loss}

\subsection*{Framework}
To identify stiff and sloppy directions, I decompose the Hessian matrix of a loss function $\mathcal{L}(\Phi, \Phi+\delta\Phi)$ that measures the change in model predictions under such a perturbation.
I use the symmetrized Kullback--Leibler divergence
$\mathcal{L}^{sKL}(\Phi,\Phi+\delta\Phi) = \tfrac{1}{2}\left[KL\!\left(P(x|\Phi)\,\|\,P(x|\Phi+\delta\Phi)\right) + KL\!\left(P(x|\Phi+\delta\Phi)\,\|\,P(x|\Phi)\right)\right]$
as the loss function given its prominence in prior work.
Throughout, I write $\Phi \equiv (\rho,\mu)^\top$ for the parameter vector and $\delta\Phi$ for an infinitesimal perturbation in log-parameter space.
Working in log-parameter space (to make derivatives scale-free and parameter-unit-agnostic), the $(i,j)$ element of the Hessian is
\begin{align}\label{eq:hessian_general}
    H_{i,j}^{sKL}(\Phi) &:= \left.\frac{d^2\mathcal{L}^{sKL}(\Phi,\Phi+\delta\Phi)}{d\!\log\Phi_{i}\,d\!\log\Phi_{j}}\right|_{\delta\Phi=0}\nonumber\\
    &= \Phi_i\Phi_j \left.\frac{d^2\mathcal{L}^{sKL}(\Phi,\Phi+\delta\Phi)}{d\Phi_i\,d\Phi_j}\right|_{\delta\Phi=0}
\end{align}
The second equality uses the fact that $\delta\Phi = 0$ is a stationary point of $\mathcal{L}^{sKL}(\Phi,\Phi+\delta\Phi)$, eliminating the chain-rule cross-term between log- and bare-parameter derivatives.
By a second-order Taylor expansion, the loss is approximated as $\mathcal{L}(\Phi, \Phi+\delta\Phi) \approx \tfrac{1}{2}\delta\Phi^\top H \delta\Phi$, so the eigenvalue-eigenvector pairs $(\lambda_i, v_i)$ of $H$ identify the principal directions and their stiffness.
The stiff direction (largest $\lambda$) induces the largest loss for a unit step; the sloppy direction (smallest $\lambda$) induces the smallest.

\subsection*{sKL Hessian for the stationary distribution of ants}
\begin{figure}[htpb]
    \centering
    \includegraphics[width=\columnwidth]{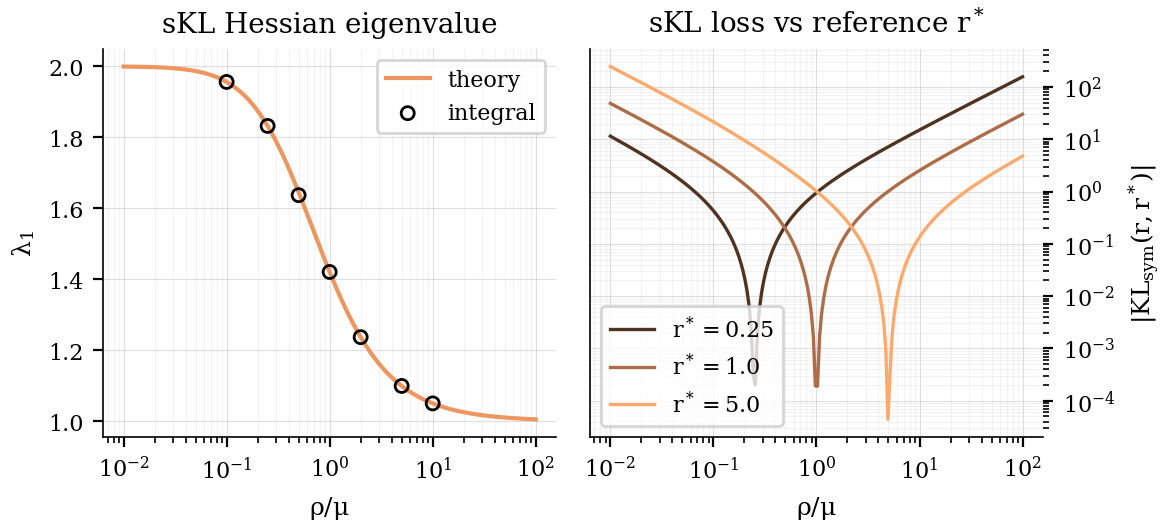}
    \caption{Left: first eigenvalue $\lambda_1$ of the sKL Hessian as a function of $\rho/\mu$. Right: the corresponding loss function evaluated at different reference points. The sKL eigenvalue decreases from $\lambda_1^{stat}\to2$ as $\rp\to0^+$ toward $\lambda_1^{stat}\to1$ as $\rp\to\infty$}
    \label{fig:hessian_theory}
\end{figure}

The Hessian of the sKL loss equals the FIM~\cite{QuinnEtAl2023InformationGeometryMultiparameter}:
\begin{equation}\label{eq:hessian_skl}
    H_{i,j}^{stat}(\Phi) = \Phi_i\Phi_j \int_0^1 P(x|\Phi)\,
        \frac{\partial\log P(x|\Phi)}{\partial\Phi_i}\,
        \frac{\partial\log P(x|\Phi)}{\partial\Phi_j}\,dx.
\end{equation}
For the Beta distribution in Eq.~\eqref{eq:ants_stationary_distribution} the FIM evaluates to
\begin{equation}\label{eq:ants_hessian_explicit_skl}
    H^{stat}(\Phi) = \frac{\rho^2}{\mu^2}
    \begin{bmatrix} 1 & -1 \\ -1 & 1 \end{bmatrix}
    \left[2\psi^{(1)}\!\left(\rp\right) - 4\psi^{(1)}\!\left(2\rp\right)\right],
\end{equation}
where $\psi^{(1)}(\cdot)$ is the trigamma function.
The eigendecomposition yields
\begin{align}
    v_1^{stat} &= \tfrac{1}{\sqrt{2}}\begin{bmatrix}-1\\1\end{bmatrix}, &
    \lambda_1^{stat} &= \frac{\rho^2}{\mu^2}
        \bigl[4\psi^{(1)}\!\left(\rp\right) - 8\psi^{(1)}\!\left(2\rp\right)\bigr], \label{eq:eval1}\\
    v_2^{stat} &= \tfrac{1}{\sqrt{2}}\begin{bmatrix}1\\1\end{bmatrix}, &
    \lambda_2^{stat} &= 0. \label{eq:eval2}
\end{align}
The first direction $v_1^{stat} \propto (-1,1)^\top$ corresponds to changes in the log-parameter ratio $\log(\rho/\mu)$, which is precisely the only parameter combination that affects the stationary distribution.
The second direction $v_2^{stat} \propto (1,1)^\top$ changes both $\log\rho$ and $\log\mu$ equally, leaving $\rho/\mu$ unchanged and hence all predictions invariant; its eigenvalue is exactly zero.
$v_2^{stat}=(1,1)$ corresponds to the global rescaling $(\rho,\mu)\mapsto(c\rho,c\mu)$, which rescales time and leaves the stationary problem invariant, so its eigenvalue is identically zero by construction.
Note that the MSE-based Hessian, as commonly used in the literature, computed on the log-variance of the Beta distribution as a representative moment, yields the same eigenvectors because the variance of $\mathrm{Beta}(\alpha,\alpha)$ depends only on $\alpha = \rho/\mu$, the same parameter combination that determines the entire distribution; the eigenvalue magnitude differs: $\lambda_1^{MSE} = 4(\rho/\mu)^2(2\rho/\mu+1)^{-2}$.
Figure~\ref{fig:hessian_theory} shows $\lambda_1(\rp)$ for the sKL loss function across the $\rp$ range.

\subsection*{sKL Hessian for the fixed-lag autocovariance}
Continuing with the sKL, the joint distribution generalizes Eq. \eqref{eq:hessian_skl} 
\begin{equation}\label{eq:hessian_skl_2D}
    H_{i,j}^{joint}(\Phi) = \Phi_i\Phi_j\int P(z|\Phi)\,
        \frac{\partial\log P(z|\Phi)}{\partial\Phi_i}\,
        \frac{\partial\log P(z|\Phi)}{\partial\Phi_j}\,dz,
\end{equation}
where the integral runs over $z=(x_t,x_{t+\tau})\in[0,1]^2$.

Using the $n=1$ slowest-mode case, whose joint density is given by Eq. \eqref{eq:joint_density_n1}, the Fisher information separates into the stationary Hessian and a rank-one term carrying the lag (see Appendix \ref{appx:joint_hessian}),
\begin{equation}\label{eq:joint_hessian}
    H^{joint}(\Phi) \approx 2H^{stat}(\Phi) + e^{-4\rho\tau}\,\beta\beta^\top,
    \qquad
    \beta = \frac{\partial\log\Delta}{\partial\log\Phi},
\end{equation}
where $\beta$ is the log-parameter gradient of the coupling $\Delta = e^{-2\rho\tau}(x_t-\tfrac12)(x_{t+\tau}-\tfrac12)/V$.
Diagonalising $H^{joint}$, the eigenvectors coincide with the stationary stiff and zero-eigenvalue directions, $(-1,1)^\top$ and $(1,1)^\top$, up to a rotation of order $e^{-4\rho\tau}$, with eigenvalues $\lambda_1^{joint}\approx 2\lambda_1^{stat}$ and $\lambda_2^{joint}\approx 2\rho^2\tau^2 e^{-4\rho\tau}$, where $\lambda_1^{stat}$ is the stationary eigenvalue of Eq. \eqref{eq:eval1}.
There is a nonzero second eigenvalue for any finite $\tau>0$, which shrinks towards zero as $\tau\to\infty$ in the $n=1$ slowest mode case (see Eq. \eqref{eq:joint_density_n1}).
In this regard, the model now admits the typical sloppy property of an eigenvalue spectrum, with the eigenvalues being separated by an order of magnitude.
See Appendix \ref{appx:joint_density_eigdecomp} for details.

\section{KDE-Based Numerical Estimation}

For general stochastic models the stationary distribution is unknown.
I estimate $P(x|\Phi)$ nonparametrically using kernel density estimation (KDE) on the simulated time series $\mathcal{X} = \{x_{s,t}\}_{s=1,\dots,S;\;t=1,\dots,T}$, this being a common method for getting an estimate of the distribution (in recent work, Mottes et al.~\cite{MottesEtAl2026GradientbasedOptimizationExact} alternatively use differentiable histograms. However, this would then depend on the binning choice).
This section details the steps of estimating $P(x|\Phi)$ given simulations $x_t(\Phi)$ for a fixed length $T$, at a sampling $dt$, and for a given random seed.

\subsection*{Logit transform for bounded state}
The state variable $x_t \in [0,1]$ is bounded, causing standard KDE to incur boundary bias.
Following~\citet{KoekemoerSwanepoel2008TransformationKernelDensity} and~\citet{WandEtAl1991TransformationsDensityEstimation}, I apply a logit transformation $g(x) = \log\frac{x}{1-x}$, mapping $x$ to $\mathbb{R}$, estimate the KDE on the transformed sample, and recover the density on $[0,1]$ via
\begin{equation}\label{eq:logit_pdf}
    \hat{P}(x|\Phi) = \frac{1}{x(1-x)}\,\hat{P}_g\!\left(g(x)|\Phi\right).
\end{equation}
Since the KL divergence is invariant under invertible transformations~\cite{CsiszarShields2004InformationTheoryStatistics}, the FIM computed in the transformed space equals that in the original space.
In the following steps I apply the KDE to the transformed data and compute the FIM directly in this space.

\subsection*{Density Estimation}
To estimate the probability density, I apply a Gaussian KDE.
Writing $\mathcal{Y}=\{y_{s,t}\}$ for the transformed sample $y_{s,t}=g(x_{s,t})$ with dimension $D$ ($D=1$ for the stationary case, and $D=2$ for the joint density).
The KDE estimator is
\begin{align}\label{eq:kde}
    \hat{P}(y|\Phi) =&~ \frac{1}{|\mathcal{Y}|} \sum_{y_i \in \mathcal{Y}} (2\pi)^{-D/2}|\mathbf{\Sigma}|^{-1/2} \exp(q)\\
    q =&~ -\tfrac{1}{2}(y-y_i)^\top \mathbf{\Sigma}^{-1}(y-y_i).\nonumber
\end{align}
where $\mathbf{\Sigma}$ is the bandwidth matrix, which I take isotropic, $\mathbf{\Sigma}=h^2\mathbf{I}_{D}$, so that a single bandwidth $h$ controls the smoothing.

Evaluation of Eq. \eqref{eq:kde} directly requires $\mathcal{O}(|\mathcal{Y}|M)$ operations for a density evaluated at $M$ points, which becomes impractical for large sample sizes. 
I thus use the binned fast-Fourier estimator \citep{Silverman1982FFT, Wand1994FastComputation} as implemented in KDEpy.
This replaces direct estimation via the sum over samples by two steps on a fixed equispaced evaluation grid $\{v_k\}_{k=1}^M$ with $M$ points, spacing $\delta$, and whose coordinates I index as $d=1,\dots,D$.
First, a unit weight is allocated to the neighbouring grid points for each sample point, split linearly by distance
\begin{align}\label{eq:binning}
    c_k = \sum_{y_i\in\mathcal{Y}} \prod_{d}\max\left(0, 1-\frac{|y_{i,d}-v_{k,d}|}{\delta}\right),
\end{align}
such that  $\sum_kc_k=|\mathcal{Y}|$ is the weight accumulated at gridpoint $v_k$ and only the $2^D$ gridpoints surrounding the sample receive weight.
The density then follows from a discrete convolution of those weights with the kernel evaluated on the same grid
\begin{align*}
    \hat{P}(v_k|\Phi) = \frac{1}{|\mathcal{Y}|}\sum_{l} c_l K_{\mathbf{\Sigma}}(v_k-v_l),
\end{align*}
with $l$ running over the same grid as $k$.
This recovers Eq. \eqref{eq:kde} up to the binning in Eq. \eqref{eq:binning}, and the convolution theorem evaluates it in $\mathcal{O}(M\log M)$ operations.

The choices to be made for estimation thus involve:
(i) the bandwidth of the estimator, $h$, which controls the tradeoff between bias and variance, with a too low bandwidth leading to sample variance and a high one to oversmoothing.
In this work I run a series of different bandwidths as a sensitivity analysis to assess how dependent the resulting eigenvector and eigenvalue estimates are to the estimator. 
For KDE fitting the bandwidth is the most sensitive parameter, while the following two decisions have less effect on the outcome. 
(ii) the kernel, which has limited effect on the results~\cite{Epanechnikov1969NonparametricEstimationMultivariate}, and I thus resolve to the simple Gaussian case.
(iii) the grid, which must extend beyond the datapoints as any mass outside the grid is lost.
Furthermore, the grid spacing $\delta$ should be small relative to the bandwidth $h$ such that the shape of the  discretised kernel is preserved.
In practice, for the log-space I choose $y\in[-18,18]$ as the bounds. 
For the grid spacing, I fix $\delta$ relative to $h$ to have at least 40 points in the discretized kernel for every bandwidth. 

\subsection*{Finite-difference FIM estimate}
Given the KDE estimate $\hat{P}(x|\Phi)$, I compute the log-parameter derivatives of $\log\hat{P}$ by central finite differences with step size $\varepsilon$ in log-space:
\begin{equation}\label{eq:fd_derivative}
    \frac{\partial\log\hat{P}(y|\Phi)}{\partial\log\Phi_i}
    \approx
    \frac{\log\hat{P}(y|\Phi e^{+\varepsilon e_i})
         - \log\hat{P}(y|\Phi e^{-\varepsilon e_i})}
         {2\varepsilon},
\end{equation}
where $e_i$ is the $i$-th unit vector.
The numerical FIM element $\hat{H}_{ij}$ is then estimated by integrating over the discretized support,
\begin{equation}\label{eq:fim_numerical}
    \hat{H}_{ij}(\Phi)
    = \int \hat{P}(y|\Phi)\,
        \frac{\partial\log\hat{P}}{\partial\log\Phi_i}
        \frac{\partial\log\hat{P}}{\partial\log\Phi_j}\,dy,
\end{equation}
where the sample $\mathcal{Y}$ pools the observations of all $S$ seeds into a single density estimate rather than averaging $S$ separate ones. Each seed is run for $T$ time units and recorded every $\delta t = 10^{-3}$, so $|\mathcal{Y}| = S\,T/\delta t$ and the budget $N = S \times T$ quoted throughout is measured in time units rather than in sample points.
The eigenvectors and eigenvalues are then obtained by SVD of the assembled $2\times 2$ matrix.

In this work, I estimate the Hessian by finite differences rather than by automatic differentiation.
I choose this approach because it treats the model as a \textit{black box} whose only requirement is that it can be run with different sets of perturbed parameters.
Thus it can be applied to models whose implementations cannot be differentiated, as is the case for many macroeconomic agent-based models that are still typically written in non-differentiable languages (e.g. C++ for \citet{DosiEtAl2010SchumpeterMeetingKeynes}, Java for \citet{CaianiEtAl2016AgentBasedStockFlow}, Matlab for \citet{CarvalhoGuilmi2020TechnologicalUnemploymentIncome}, or NetLogo) and contain a variety of discrete choices that may break the pathwise derivative, even before it would be converted to a density.
Where a differentiable implementation is available, the score can instead be obtained pathwise by differentiating the density estimator through the simulated sample. 
For instance, \citet{MottesEtAl2026GradientbasedOptimizationExact} recently applied this with differentiable histogram and using a Gumbel-Softmax trick for differentiating through discrete choices.

There are also alternative approaches to estimating the density of the function, including simulation based and variational methods \citep[e.g. see][]{CranmerEtAl2020FrontierSimulationbasedInference,BleiEtAl2017VariationalInferenceReview}, however, at the cost of introducing a trained model between the simulator and the model, with a larger number of hyper-parameters to tune. 
By contrast, KDE based estimation of distributions is a standard approach with few tunable parameters, that being primarily the bandwidth, for which a sensitivity analysis is easily done once all numerical differences are run.

\section{Convergence of the KDE-Based Hessian}
This section first highlights the convergence properties of the estimation pipeline for the simpler stationary distribution.
Second, the section on the joint density extends this to assess the case of a second non-zero eigenvalue as a function of the autocovariance lag $\tau$.

\subsection*{The stationary 1D case}
This section presents the results of estimating the FIM for the stationary one-dimensional distribution, with two distinct findings: 
First, the eigenvectors, which fix the geometry of the loss surface, are estimated accurately at low data budgets and are insensitive to the choice of bandwidth in the KDE.
Second, the associated eigenvalues, which encode the scale, converge slowly and depend strongly on the choice of bandwidth. 
The remainder of the section discusses these effects in order.

\begin{figure}[htpb]
    \centering
    \includegraphics[width=\columnwidth]{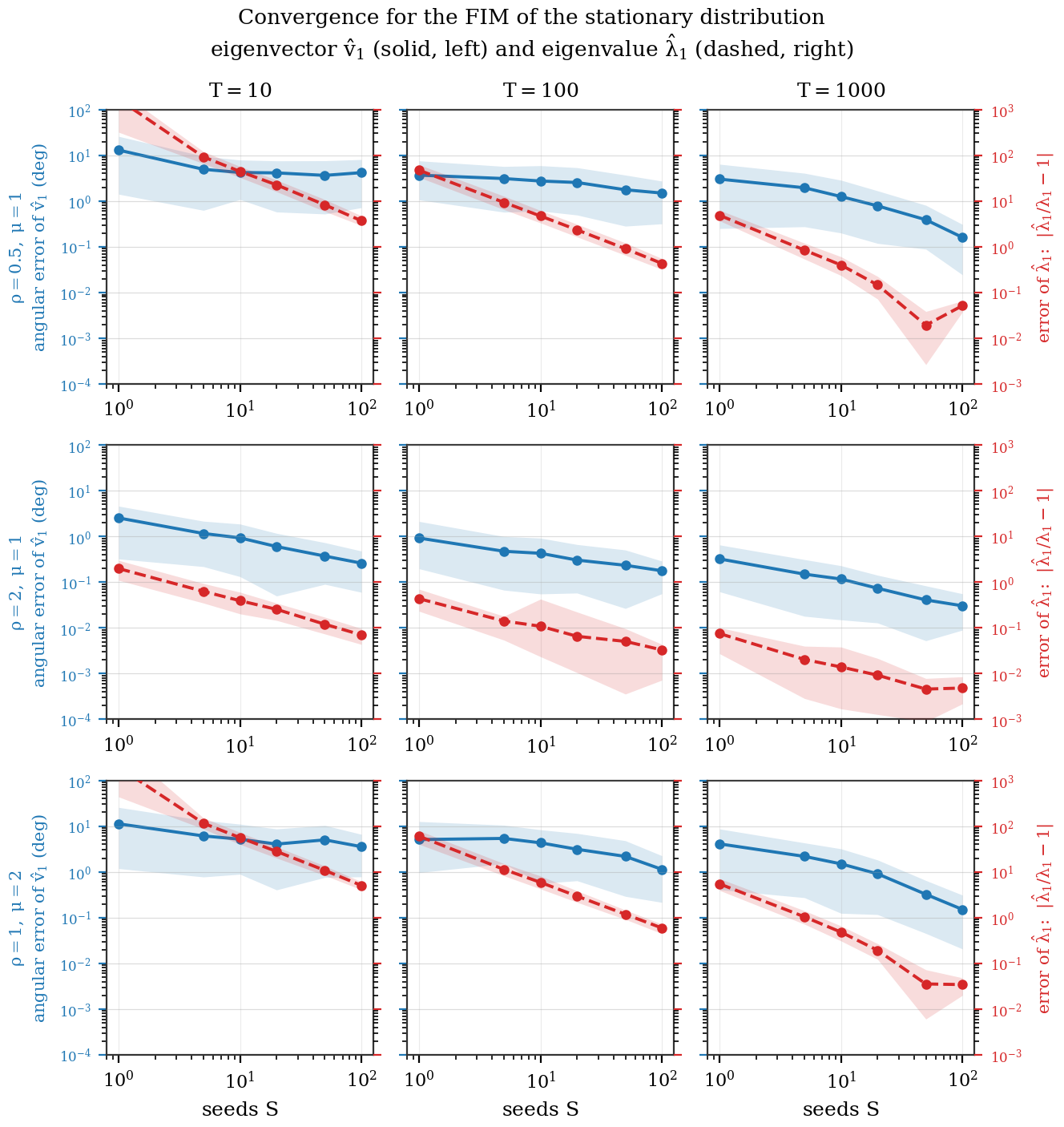}
    \caption{Convergence of the estimated Fisher information of the stationary distribution, across seeds and timesteps.
        Columns give the run length $T$ and the horizontal axis the number of seeds $S$; rows give the three parameter points, two of which share $\rho/\mu = 0.5$ and differ only in the relaxation rate.
        The solid blue curve and left axis give the angular error of the stiff eigenvector $\hat{v}_1$ against the analytical $v_1^{stat} = (-1,1)^\top/\sqrt{2}$.
        The dashed red curve and right axis give the relative error of the stiff eigenvalue, $|\hat{\lambda}_1/\lambda_1 - 1|$, taken per subset before averaging so that over- and under-estimates cannot cancel.
        Curves are means over 100 resampled seed subsets and bands span the tenth to ninetieth percentile.
    }
    \label{fig:convergence}
\end{figure}

Figure~\ref{fig:convergence} shows the convergence of the estimated first (stiff) eigenvector $\hat{v}_1$ and its associated eigenvalue $\hat{\lambda}_1$ to its analytical counterparts as a function of the number of simulation seeds $S$ and simulation lengths $T$ (Figures \ref{fig:eigenvector_convergence} and \ref{fig:eigenvalue_convergence} in Appendix \ref{appx:convergence} show boxplots of the convergence of eigenvector and eigenvalues respectively), while keeping the bandwidth fixed as $h=0.1$.
The quality of fit for the eigenvector is measured by the angular error relative to the theoretical target, and the fit of the eigenvalue by the excess $\hat{\lambda}_1/\lambda_1-1$ relative to the theoretical value.

While both converge, the convergence rate is very different, with the eigenvectors converging rapidly relative to the eigenvalues.
At $S=10$ and $T=100$ the estimated direction is already within $2.7^\circ$ at the bimodal point, while the eigenvalue is too large by a factor of six.
At the largest budget the direction is accurate to $0.16^\circ$ and the eigenvalue to $5\%$.
The unimodal point converges faster and to a higher degree of accuracy, reaching $0.03^\circ$ and $0.5\%$.

Increasing the budget by seeds or by run length is close to, but not exactly, equivalent.
The slowest relaxation mode sets an ergodic time $\tau_R = (2\rho)^{-1}$ \citep{MoranEtAl2020SchrodingersAntsContinuous}, and every run length tested exceeds it by at least an order of magnitude.
A residual run-length effect nonetheless remains.
At $T=10$ the angular error saturates near $4^\circ$ and a twenty-fold increase in seeds improves it by only a factor of $1.2$, whereas at $T=1000$ the same increase improves it by a factor of $12$.
The two points sharing $\rho/\mu = 0.5$ isolate this: they have identical tails but ergodic times of $1.0$ and $0.5$, and the point that fits more relaxation times into a run saturates less.
Quasi-non-ergodicity \citep{BouchaudFarmer2023SelfFulfillingProphecies} is therefore not the binding constraint at these budgets, but run length is not entirely interchangeable with seed count either.

\begin{figure}[htpb]
    \centering
    \includegraphics[width=\columnwidth]{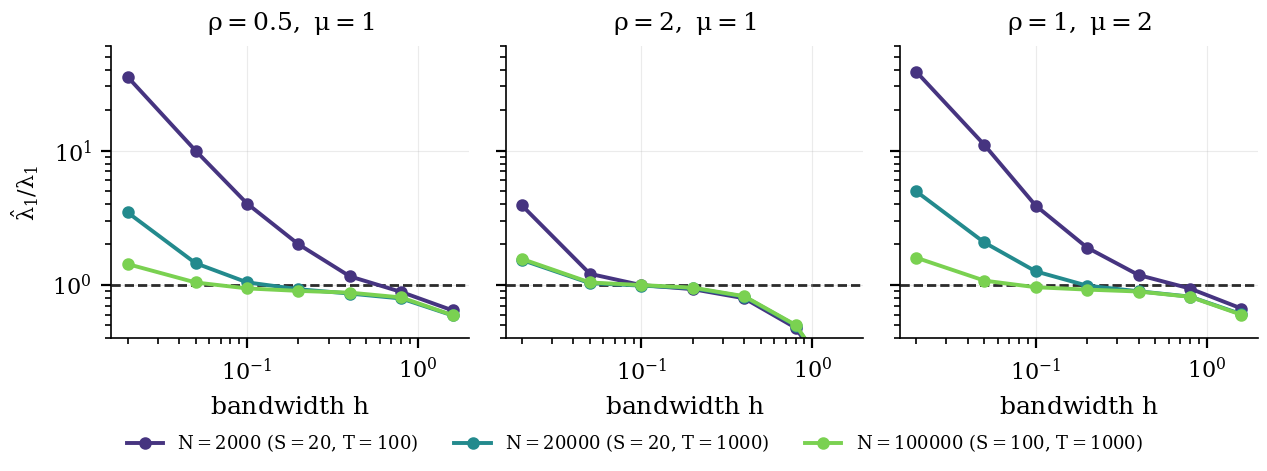}
    \caption{Stiff eigenvalue against the KDE bandwidth $h$, at three simulation budgets $N = S \times T$ spanning a fifty-fold range, for the three parameter points.
        The dashed line marks the analytical value.}
    \label{fig:bandwidth_scan}
\end{figure}

Convergence in the bimodal case is slower than the unimodal regime, reflecting the heavier tails of the Beta distribution near the boundaries, which require more data for the logit-transformed KDE to resolve accurately.
The information that identifies the eigenvalue sits in the tails, where the sample count falls exponentially. 
While the reporting grid in this case extends far enough, the fit of the tails remains a balancing act as a wide KDE bandwidth smooths away the slope carrying the signal, introducing a systematic bias, while a narrow small bandwidth increases the variance of the estimand. 
This explains the direction of the eigenvalue error for small bandwidths, as the underresolved tails lead to a gradient of the tails that is too steep, which inflates the FIM. 
Figure~\ref{fig:bandwidth_scan} separates the two effects by comparing different bandwidth at three simulation budgets.
At small $h$ the curves are spread over multiple orders of magnitude, with the error dominated by sampling noise and thus falling sharply with the budget. 
The consequence being that even for a high simulation budget, the estimate is above the theoretical value.
At large $h$ the three data curves collapse onto a common curve, implying that the error is a smoothing bias that increased data doesn't remove.
This can be seen particularly for bandwidths above 0.4, where the estimates are below the theoretical value and similar irrespective of the simulation budget.
From a practical point of view, without prior knowledge of the distribution, once the simulations are run one can select a bandwidth where there is still some data-led movement, just before the lines overlap and one is likely in a biased regime. Here, in the unimodal case this would correspond to about 0.1, or slightly below.
The three points cross the analytical value near $h \approx 0.1$, which is the bandwidth used for Fig \ref{fig:convergence}.

The simple estimation process I propose here has no prior knowledge of the rank deficiency in the stationary Hessian, which has $\lambda_2^{stat}=0$. 
In the simulations, all estimates return a positive second eigenvalue with a minimum of $2.4\times10^{-5}$ at the largest data budget.
This small but positive quantity is the effect of sampling and smoothing. 
At a budget of $N=100$ the ratio $\hat\lambda_2/\hat\lambda_1$ is $0.18$, $0.008$ and $0.24$ at the three parameter points, falling to $0.0086$, $0.000063$ and $0.011$ at the largest.
Thus, the estimator approaches the degeneracy without being told that it exists.

These first results suggest that a standard Gaussian KDE pipeline using finite difference can recover the theoretical FIM eigendecomposition of Kirman's ant model without requiring knowledge of the analytical distribution. 
In particular, with only moderate data requirements one achieves good estimates of the geometry of the loss surface, with eigenvector directions converging quickly. 
By contrast, the scale requires significantly more data due to the tails of the Beta distribution, and is sensitive to the choice in bandwidth.\footnote{Previous versions of this work applied the Improved Sheather--Jones bandwidth. In this context, the ISJ returns $h$ between $0.02$ and $0.07$, which lies in the noise-dominated regime of Figure~\ref{fig:bandwidth_scan}, and inflates $\hat\lambda_1$ by between $46\%$ and $136\%$ at the three parameter points with $N=20000$ observations. The reason is that the selector optimises the integrated squared error of the
density, whereas the Fisher integral weights the tails, where the density carries little mass.}

\subsection*{The joint density case}
\begin{figure}[htpb]
    \centering
    \includegraphics[width=\columnwidth]{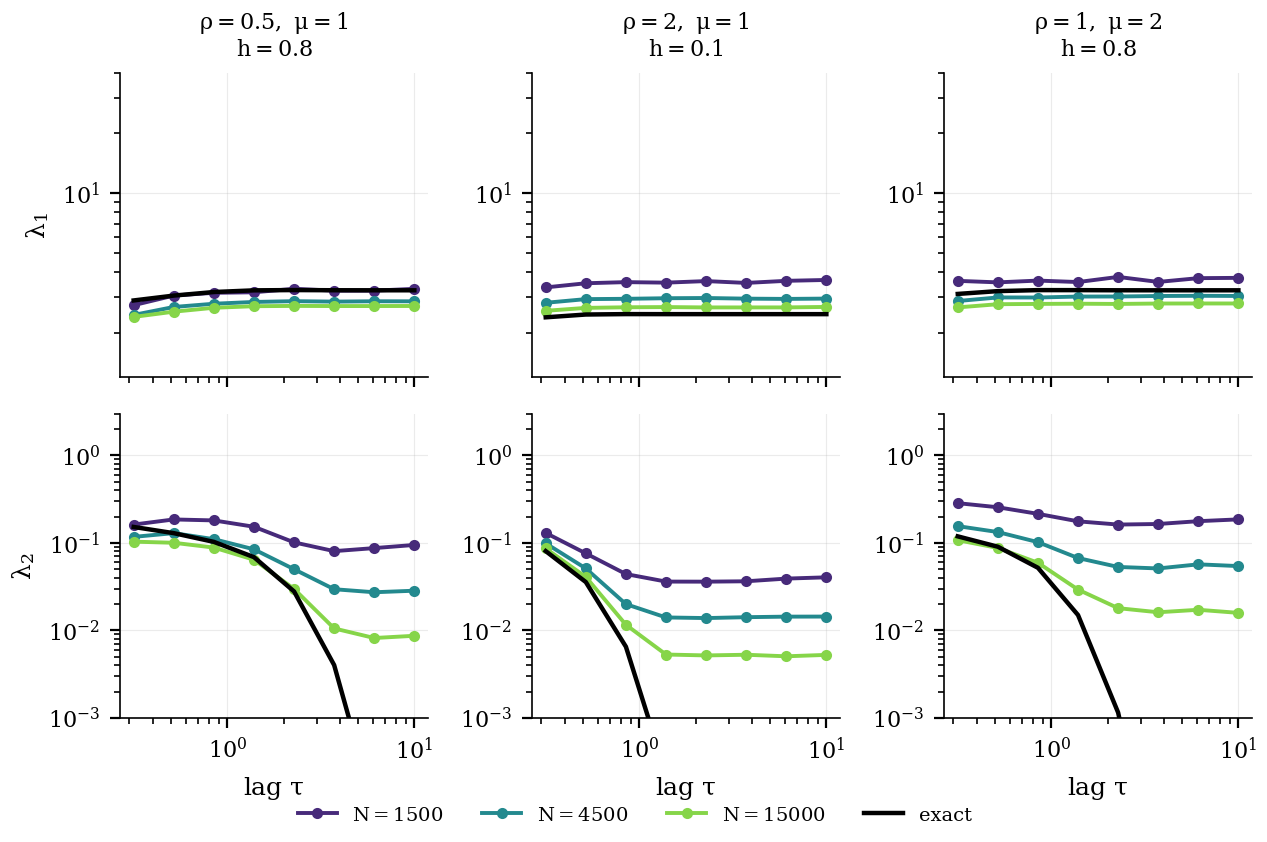}
    \caption{Eigenvalues of the fixed-lag sKL Hessian against the lag $\tau$, at three parameter points.
        The stiff eigenvalue $\lambda_1$ is in the top row and the sloppy eigenvalue $\lambda_2$ in the bottom row.
        Black gives the exact spectral result, which sums all relaxation modes.
    Coloured lines give the KDE estimate at the fixed bandwidth $h$ named in each panel, for three total sample lengths $N = S \times T$ built from $S = 15$ seeds.}
    \label{fig:joint_eigenvalues}
\end{figure}

The exact zero second eigenvalue in the stationary spectrum corresponds to a rescaling of time, leaving the stationary outcome unchanged.
By contrast, the joint density $\langle x_t,x_{t+\tau}\rangle$ breaks that symmetry, as the relaxation of the autocovariance depends on $\rho$. 
This leads to a Hessian that is full-rank and a non-zero second eigenvalue, admitting a typical sloppy spectrum.
Figure~\ref{fig:joint_eigenvalues} reports the two eigenvalues of the resulting Hessian as a function of the lag $\tau$, comparing different data-quantities $N=S\cdot T$ (lines) to the exact theoretical eigenvalue for a parameter-specific bandwidth (solid black line).
The bandwidth shown here is selected as the best fitting candidate out of 5 options based on the criteria in Section~V.A, to isolate the effect of increased data rather than changes in bandwidth. 
The fit of different bandwidths is shown in Fig \ref{fig:joint_bandwidth} of Appendix \ref{appx:joint_bandwidth}, here varied across the lag rather than the number of datapoints.

The stiff eigenvalue is recovered accurately, with close estimates already at low simulation budgets.
In the bimodal case, the estimate drifts below the exact value for the first eigenvalue as the simulation budget grows, revealing the smoothing bias in the KDE estimator.
The sloppy eigenvalue is small and strictly positive, which is the behaviour the stationary case cannot exhibit.
It decreases with the lag and follows the exact spectral result over roughly a decade in $\tau$.
At this point, the estimates plateau, with the value of the plateau dropping as a function of the simulation budget.
Once the correlation $e^{-2\rho\tau}$ has decayed the joint density factorises into the product of its marginals, and the Hessian becomes rank one.
Any second eigenvalue reported at large lag therefore measures the sampling noise that breaks the exact product of marginals structure.
It falls by about one order of magnitude as the total sample size grows by an order of magnitude.

\begin{figure}[htpb]
    \centering
    \includegraphics[width=\columnwidth]{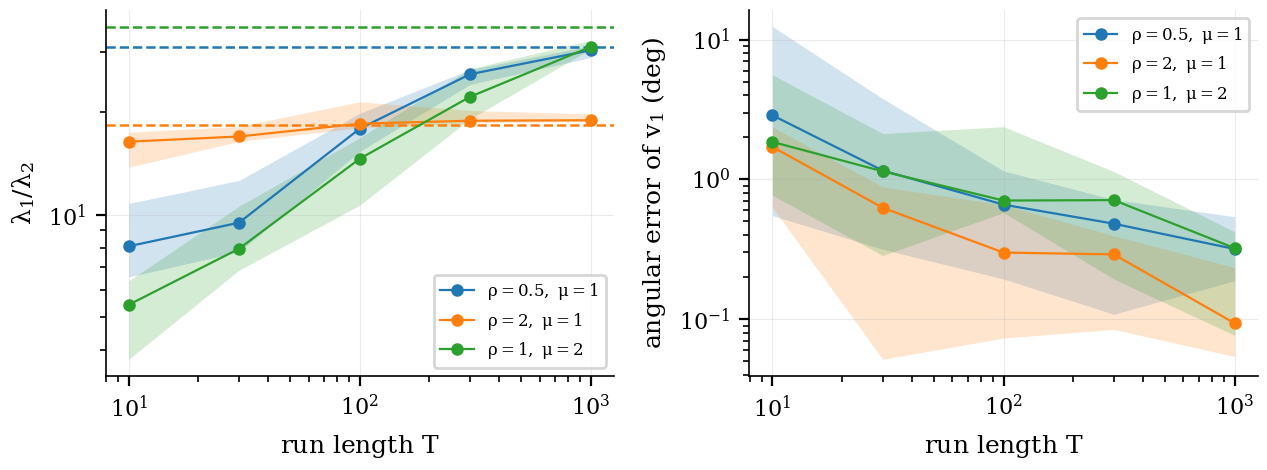}
    \caption{Convergence of the fixed-lag estimate in the run length $T$, evaluated at a lag of order the relaxation time, where $\lambda_2$ sits clear of the estimator plateau of Fig.~\ref{fig:joint_eigenvalues}.
        Left: the condition number $\lambda_1/\lambda_2$, which measures sloppiness, with the theoretical value as a dashed horizontal line.
        Right: the angular error of the estimated stiff eigenvector against the exact joint eigenvector, which is itself rotated slightly away from the stationary $(-1,1)^\top/\sqrt{2}$.
        Markers give the median over ten disjoint groups of fifteen seeds, and the band gives the tenth to ninetieth percentile across those groups.
        Each point uses $15\,T$ time units of simulation.}
    \label{fig:joint_convergence}
\end{figure}

Figure~\ref{fig:joint_convergence} examines the estimate at a lag of order the relaxation time.
The condition number $\lambda_1/\lambda_2$ measures the sloppiness of the model and is recovered to within $2\%$, $3\%$ and $13\%$ of the analytical value at the three points at the largest simulation budget.
At two of the three it is more accurate than either eigenvalue separately, since both are biased low by similar factors at the bandwidth used and the biases partly cancel in the ratio.
The eigenvectors are recovered to a third of a degree, measured against the exact joint basis rather than the stationary one, since the joint eigenvectors are themselves rotated away from $(-1,1)^\top/\sqrt{2}$ by an angle of order $e^{-4\rho\tau}$.
In fashion similar to the stationary one-dimensional case, the geometry is accurately recovered before the scale.

\section{Discussion and Conclusion}

I have demonstrated that a standard KDE and finite difference pipeline is sufficient to recover the analytical FIM eigenvectors and eigenvalues of Kirman's ant model, both for its stationary distribution and for a joint-distribution.
The sKL Hessian (FIM) of the stationary distribution has a single non-zero eigenvalue, with eigenvector $v_1^{stat} = (-1,1)^\top/\sqrt{2}$ corresponding to the log-ratio $\log(\rho/\mu)$, the only parameter combination affecting the stationary distribution.
The KDE-based estimate converges to the analytical ground truth with modest simulation budgets.
Going beyond the stationary distribution, to a joint distribution of the autocovariance with a fixed lag, the FIM becomes full-rank. 
The eigenvectors are closely aligned to those of the stationary distribution, but with a non-zero second eigenvalue. 
In this context, the classical sloppy spectrum emerges, where the eigenvalues are separated by an order of magnitude. 
The simple pipeline of this paper consistently recovers the first eigenvector at a low data budget, and the first eigenvalue at a larger one. 
The second eigenvalue is the most demanding quantity, since it is the smallest and is the first to be lost in the estimator's noise floor. 

The stiff direction also suggests a way to explore a model's parameter space, along the lines of \citet{Naumann-WoleskeEtAl2025ExplorationParameterSpace}.
A second-order expansion of the loss about $\Phi$ would imply that to change the loss by a given magnitude one can follow the direction of eigenvector $i$ with a step-size proportional to $\lambda_i^{-\frac12}$. 
In the Kirman's ants model it would, if one naively starts in the bimodal case, push the parameters towards the unimodal case, thus discovering different behaviours of the model.  
While Kirman's ants allows for analytical solutions, this is not the case for larger macroeconomic agent-based models where this approach could be used to discover qualitatively different model regimes, especially as the use of outcome distributions is likely less sensitive to realization noise than the time-series based mean-squared error (non-linear least squares) used in a large share of the sloppy-models literature. 
This approach thus offers a means to estimate the relevant parameter-space directions for a principled alternative to exhaustive grid search or surrogate modeling~\cite{tenBroekeEtAl2021UseSurrogateModels,LampertiEtAl2018AgentBasedModelCalibration}.

This work is a first study with a model of only two parameters with convenient analytical properties, and it offers multiple questions for further research. 
A first question is on scaling. 
Specifically, how accurate this pipeline would be for models with $P>2$ parameters and multiple different observables.
The practical limit is set by the dimension of the observable rather than by the number of parameters.
Parameter count enters linearly, since a central-difference Jacobian requires $2P$ simulations, and those runs are independent (thus trivially parallelizable).
The density estimate is the constraint, since kernel density estimation converges more slowly as the dimension of the observable grows.
The fixed-lag case here is a two-dimensional estimate and recovers the eigenstructure at $N = 15000$ time units, which suggests that observables of three or four dimensions remain feasible while full joint densities beyond that would require a projection or a marginal.
Automatic differentiation, with exact gradients, would be an interesting extension in this regard. 
One concern could be with increases in the number of observables that could lead to distributions similar to the bimodal case with exponential tails where a prohibitive number of realizations may be necessary to accurately estimate eigenvalues.
Nonetheless, as the results suggest, geometry may be well-estimated even at low data budget, such that for activities like the exploration of the parameter space this may be sufficient, while activities requiring all eigenvalues to be well-estimated may not be feasible. 
Another question in a similar vein is on finite effects in agent-based models. 
In this paper, I used the large agent limit for Kirman's ants to compare estimand with theory. 
Given the results of this paper, one could estimate the scale of the finite effects on such estimate using the recent work of \citet{HolehouseMoran2022ExactTimedependentDynamics} who solved the exact time-dependent dynamics for this model. 
Extension to the higher-dimensional parameter spaces and joint observables typical of full agent-based models is left to future work, also to test further the geometric decay of eigenvalues.

\begin{acknowledgments}
    I thank Michael Benzaquen and Jean-Philippe Bouchaud for many discussions and their detailed comments on this work. 
    I also thank James Sethna for comments on an earlier version of the manuscript. 
    Finally, I am very grateful to the two anonymous reviewers for their comments and suggestions, in particular for the suggesting the extension to considering the finite-lag autocovariance as an example of a non-zero second eigenvalue, the discussion around potential automatic differentiation approaches, and the clarifications and corrections suggested.
    I acknowledge support from the ESG-UPTAKE project (Grant Agreement ID: 101145727).
    This research was funded in whole or in part by the Austrian Science Fund (FWF) [10.55776/ESP4179425].
\end{acknowledgments}

\bibliography{refs}

\appendix

\section{Full convergence distributions for the stationary case}\label{appx:convergence}
Figure~\ref{fig:convergence} reports means over the 100 resampled seed subsets.
The two figures below keep the full distribution behind each of those means.

\begin{figure}[htpb]
    \centering
    \includegraphics[width=\columnwidth]{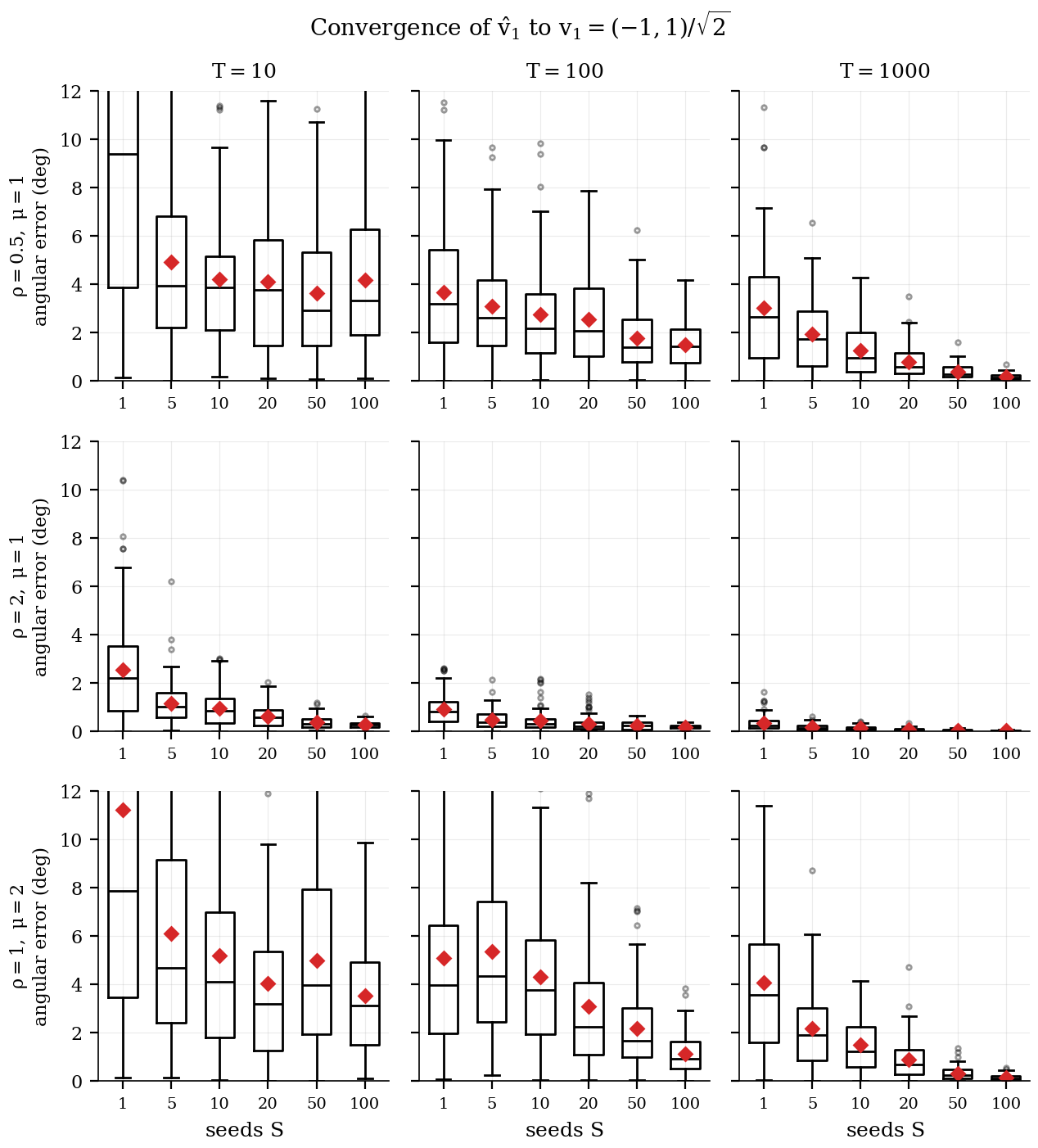}
    \caption{Angular error of the estimated stiff eigenvector $\hat{v}_1$ against the analytical $v_1^{stat} = (-1,1)^\top/\sqrt{2}$.
        Columns give the run length $T$, the horizontal axis the number of seeds $S$, and rows the three parameter points.
    Boxes span the interquartile range over 100 resampled seed subsets, with medians in black and means as red diamonds.}
    \label{fig:eigenvector_convergence}
\end{figure}

\begin{figure}[htpb]
    \centering
    \includegraphics[width=\columnwidth]{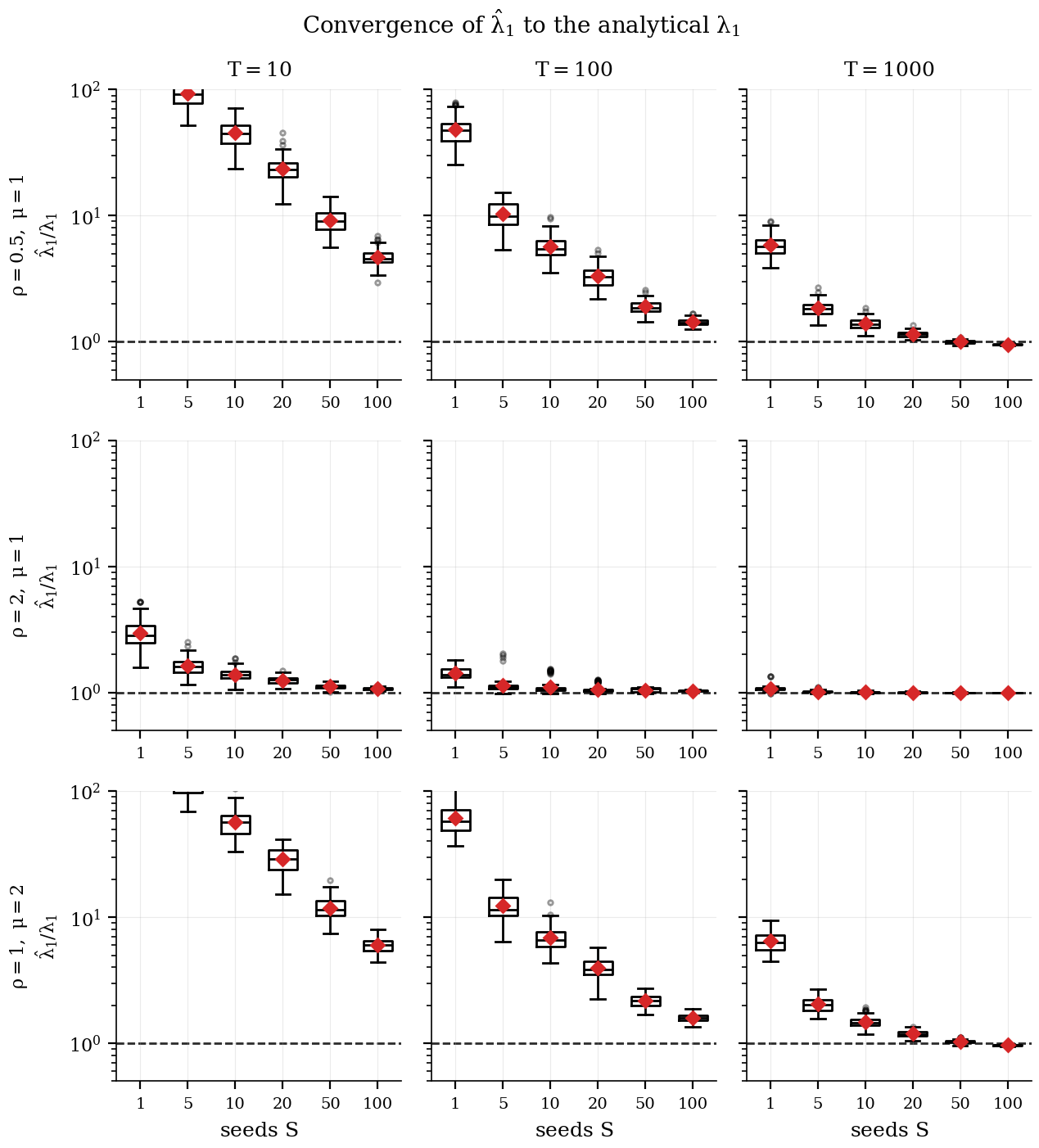}
    \caption{The ratio $\hat{\lambda}_1/\lambda_1^{stat}$. 
        Columns give the run length $T$, the horizontal axis the number of seeds $S$, and rows the three parameter points.
    Boxes span the interquartile range over 100 resampled seed subsets, with medians in black and means as red diamonds.
    The dashed line represents the target ratio of 1.0
}
    \label{fig:eigenvalue_convergence}
\end{figure}
\section{Derivation of the joint density $P(x_t,x_{t+\tau}|\Phi)$}\label{appx:joint_density_n1}

This appendix presents the steps for the simplification of the joint density $P(x_t,x_{t+\tau})$ for the case of only the slowest mode $n=1$.
Notationally, I omit here that its conditional on the point in parameter space $\Phi$.
The starting point is 
\begin{align*}
    P(x_t, x_{t+\tau}) =&~ P(x_{t})P(x_{t+\tau})\left[1 + \frac{e^{-2\rho\tau}\Psi_1(x_t)f_1(x_{t+\tau})}{P(x_{t+\tau})}\right]\label{eq:appx_joint_1_general}
\end{align*}
Define $a(x_t)=\Psi_1(x_t)$ and $b(x_{t+\tau})=\frac{f_1(x_{t+\tau})}{P(x_{t+\tau})}$.
The marginal densities imply
\begin{align*}
    P(x_t) =&~ \int P(x_t, x_{t+\tau})dx_{t+\tau}\\ 
    =&~ P(x_{t})\left[1 + e^{-2\rho \tau}a(x_t)\int b(x_{t+\tau})P(x_{t+\tau})dx_{t+\tau}\right]
\end{align*}
which requires $a(x_t)$ and $b(x_{t+\tau})$ to have zero mean
\begin{align*}
    \int b(x_{t+\tau})P(x_{t+\tau})dx_{t+\tau} =&~ 0\\
    \int a(x_{t})P(x_{t})dx_{t} =&~ 0,
\end{align*}
where the second line is the same case but for the $x_{t+\tau}$ marginal distribution.
The next observation is that Kirman's ants is a symmetric model, such that 
\begin{align*}
    P(x_t, x_{t+\tau}) =&~ P(x_{t+\tau}, x_t)
\end{align*}
which requires
\begin{align*}
    \frac{a(x_t)}{b(x_t)} =&~ \frac{a(x_{t+\tau})}{b(x_{t+\tau})} = constant
\end{align*}
which leads me to introduce the simpler $\varphi(x_t)=a(x_t)=b(x_t)$.
The joint distribution is therefore of the form
\begin{align*}
    P(x_t, x_{t+\tau}) =&~ P(x_{t})P(x_{t+\tau})\left[1 + e^{-2\rho \tau}\varphi(x_t)\varphi(x_{t+\tau})\right]
\end{align*}
Next, define $m(\tau)=\mathbb{E}(x_{t+\tau}|x_t)$ as the quantity of interest.
Returning to the SDE of Kirman's ants, the noise increment $dW_t$ has an expected value of zero by definition, which leaves just the drift term $\rho(1-2x)$.
Since the drift is linear in $x$ the average of the drift is the drift of the average,
\begin{align*}
    dm = \rho(1 - 2m)d\tau = -2\rho(m - \frac12),
\end{align*}
which has a fixed point at $m=\frac12$. 
This is an ordinary first order ODE, where $m(0)=x_t$, such that
\begin{align*}
m(\tau) = \mathbb{E}(x_{t+\tau}|x_t) = \frac12 + (x_t - \frac12)e^{-2\rho\tau}
\end{align*}

The conditional expectation $\mathbb{E}(x_{t+\tau}-\frac12|x_t)$ based on the joint density.
First, noting 
\begin{align*}
    P(x_{t+\tau}|x_t) =&~ \frac{P(x_t,x_{t+\tau})}{P(x_t)}\\
    =&~ P(x_{t+\tau})\left[1 + e^{-2\rho \tau}\varphi(x_t)\varphi(x_{t+\tau})\right]
\end{align*}
thus
\begin{align*}
    \mathbb{E}(x_{t+\tau}-\frac12|x_t) =&~ \int(x_{t+\tau}-\frac12)P(x_{t+\tau})dx_{t+\tau}\\
    ~&+ e^{-2\rho \tau}\varphi(x_t)\int(x_{t+\tau}-\frac12)\varphi(x_{t+\tau})P(x_{t+\tau})dx_{t+\tau}\\
    =&~ e^{-2\rho \tau}\varphi(x_t)c
\end{align*}
where the first integral drops as $\mathbb{E}(x_{t+\tau})=\frac12$ and I use $c$ as shorthand for the second integral, as it evaluates to a constant.

Putting together the two ways of computing the conditional expectation, 
\begin{align*}
    \mathbb{E}(x_{t+\tau}-\frac12|x_t) =&~ (x_t - \frac12)e^{-2\rho\tau} = e^{-2\rho \tau}\varphi(x_t)c\\
    \varphi(x_t) =&~ \frac{x_t - \frac12}{c}
\end{align*}
where $c$ can be determined using its definition
\begin{align*}
    c =&~ \int(x_{t+\tau}-\frac12)\varphi(x_{t+\tau})P(x_{t+\tau})dx_{t+\tau}\\
    =&~ \frac{1}{c}\int(x_{t+\tau}-\frac12)^2P(x_{t+\tau})dx_{t+\tau}\\
    =& \sqrt{Var(x)} = \sqrt{\frac{\mu}{4(2\rho+\mu)}}
\end{align*}
where the remaining integral is the variance of the stationary $\mathrm{Beta}(\rp,\rp)$ given by $V=\frac{\mu}{4(2\rho+\mu)}$

Combining this, the joint density conditional on $n=1$ can be expressed as
\begin{align*}
    P(x_t, x_{t+\tau}) =&~ P(x_{t})P(x_{t+\tau})\left[1 + e^{-2\rho \tau}\frac{(x_t-\frac12)(x_{t+\tau}-\frac12)}{V}\right]
\end{align*}

\section{sKL Hessian matrix of the joint density $P(x_t,x_{t+\tau}|\Phi)$}\label{appx:joint_hessian}
The Hessian is the FIM, which can be written as
\begin{align*}
    H_{ij} = \mathbb{E}_P\left[\frac{\partial \log P}{\partial \log \Phi_i}\frac{\partial \log P}{\partial \log \Phi_j}\right]
\end{align*}

For this section I simplify the notation, such that $P(x_t|\Phi)=P_0$ and $P(x_{t+\tau}|\Phi)=P_\tau$

Using the $n=1$ slowest-mode case, whose joint density is given by Eq. \eqref{eq:joint_density_n1}, the log-density splits into three components
\begin{align*}
    \log P(x_t, x_{t+\tau}) =&~ \log P(x_t) + \log P(x_{t+\tau}) \\ 
    &+ \log\left[1 + e^{-2\rho \tau}\frac{(x_t-\frac12)(x_{t+\tau}-\frac12)}{V}\right] 
\end{align*}
The derivatives of the stationary distributions is straightforward, 
\begin{align*}
    \frac{\partial\log P(x)}{\partial\log\rho}&=+\frac{\rho}{\mu}\Bigl[2\psi\bigl(\tfrac{2\rho}{\mu}\bigr)-2\psi\bigl(\tfrac{\rho}{\mu}\bigr)+\log\bigl(x(1-x)\bigr)\Bigr],\\
    \frac{\partial\log P(x)}{\partial\log\mu}&=-\frac{\rho}{\mu}\Bigl[2\psi\bigl(\tfrac{2\rho}{\mu}\bigr)-2\psi\bigl(\tfrac{\rho}{\mu}\bigr)+\log\bigl(x(1-x)\bigr)\Bigr],
\end{align*}
where $\psi$ is the digamma function. 

Turning to the third component, and defining $\Delta:=e^{-2\rho\tau}(x_t-\tfrac12)(x_{t+\tau}-\tfrac12)/V$, the derivative of the third part is
\begin{align*}
    \frac{\partial\log(1+\Delta)}{\partial\log\Phi_i}=\frac{1}{1+\Delta}\frac{\partial\Delta}{\partial\log\Phi_i}=\frac{\Delta}{1+\Delta}\frac{\partial\log\Delta}{\partial\log\Phi_i}
\end{align*}
where
\begin{align*}
\frac{\partial\log\Delta}{\partial\log\Phi_i} =&~\frac{\partial}{\partial\log\Phi_i}\left[-2\rho\tau - \log V\right]\\ 
=&~\frac{\partial}{\partial\log\Phi_i}\left[-2\rho\tau - \log\mu +\log(2\rho+\mu)\right] 
\end{align*}
where all constants with zero derivative for $\Phi_i$ are dropped (e.g. the terms in $x_t$).
The $\rho$ and $\mu$ log-derivatives are
\begin{align*}
\frac{\partial\log\Delta}{\partial\log\rho} =&~\frac{2\rho}{2\rho + \mu} - 2\rho\tau\\
\frac{\partial\log\Delta}{\partial\log\mu} =&~-\frac{2\rho}{2\rho + \mu}
\end{align*}
with the common component comes from the variance term.

At this point, I can now construct $H_{ij}^{joint}(\Phi,\tau)$ for the joint density. 
There are three components to the product in $\mathbb{E}_P\left[\frac{\partial \log P}{\partial \log \Phi_i}\frac{\partial \log P}{\partial \log \Phi_j}\right]$:
\begin{enumerate}
    \item The product of the stationary components.
    Let $s=\log P(x_t|\Phi) + \log P(x_{t+\tau}|\Phi)$ and one finds
    \begin{align*}
        \mathbb{E}\left[\frac{\partial s}{\partial \log \Phi_i}\frac{\partial s}{\partial \log \Phi_j}\right] = 2 H^{stat}_{ij}
    \end{align*}
    where the same-time products each average to the $H^{stat}_{ij}$ of Eq. \eqref{eq:hessian_skl} and the two cross-time products are zero.
    \item The product of the coupling term 
    \begin{align*}
        ~&~\mathbb{E}\left[\frac{\partial \log(1+\Delta)}{\partial \log \Phi_i}\frac{\partial \log(1+\Delta)}{\partial \log \Phi_j}\right]\\
        =&~\mathbb{E}\left[\left(\frac{\Delta}{1+\Delta}\right)^2\right]\frac{\partial \log\Delta}{\partial \log \Phi_i}\frac{\partial \log\Delta}{\partial \log \Phi_j}\\
        =&~I(\tau)\frac{\partial \log\Delta}{\partial \log \Phi_i}\frac{\partial \log\Delta}{\partial \log \Phi_j}
    \end{align*}
    where 
    \begin{align*}
        I(\tau) =&~ \mathbb{E}\left[\left(\frac{\Delta}{1+\Delta}\right)^2\right]\\
        =&~ \iint \frac{\Delta^2}{1+\Delta}P(x_t)P(x_{t+\tau})dx_tdx_{t+\tau}\\
        =&~ \sum_{m\geq2}(-1)^m\iint\Delta^mP(x_t)P(x_{t+\tau})dx_tdx_{t+\tau}\\
        =&~ \sum_{m\geq2}(-1)^m \frac{e^{-2m\rho\tau}}{V^m}\left[\int\left(x-\frac12\right)^mP(x)dx\right]^2
    \end{align*}
    for $|\Delta|<1$ such that I can expand $\frac{\Delta^2}{1+\Delta}$ as a geometric series. For odd $m$ the integral term is zero as $x-\frac12$ is odd about $\frac12$ and $P$ is symmetric. Thus only even $m$ remain, such that I can approximate with $m=2$ and an error term of order $e^{-8\rho\tau}$
    \begin{align*}
        I(\tau) =&~ e^{-4\rho\tau} + O(e^{-8\rho\tau})
    \end{align*}
    where for $m=2$ the integral term cancels the variance term.
    \item The cross-term 
    \begin{align*}
        \mathbb{E}\left[\frac{\partial s}{\partial \log \Phi_i}\frac{\partial \log (1+\Delta)}{\partial \log \Phi_j}\right] = 0
    \end{align*}
    because
    \begin{align*}
        &\iint P(x_t)P(x_{t+\tau})\frac{\partial s}{\partial\log\Phi_i}\frac{\partial\Delta}{\partial\log\Phi_j}dx_tdx_{t+\tau}\\
        &=c\iint P(x_t)P(x_{t+\tau})\frac{\partial s}{\partial\log\Phi_i}\left(x_t-\frac12\right)\left(x_{t+\tau}-\frac12\right)dx_tdx_{t+\tau}
    \end{align*}
    with $c=\frac{\partial\log\Delta}{\partial\log\Phi_j}\frac{e^{-2\rho\tau}}{V}$. The double integral can be expanded and split into a product of single integrals
    \begin{align*}
        &~\left[\int \frac{\partial \log P(x_t)}{\partial\log\Phi_i}\left(x_t-\frac12\right)P(x_t)dx_t\right]\underbrace{\left(\mathbb{E}(x_{t+\tau})-\frac12\right)}_{=0}\\
        &+\underbrace{\left(\mathbb{E}(x_{t})-\frac12\right)}_{=0}\left[\int \frac{\partial \log P(x_{t+\tau})}{\partial\log\Phi_i}\left(x_{t+\tau}-\frac12\right)P(x_{t+\tau})dx_{t+\tau}\right]
    \end{align*}
    where the zero mean property leads the cross-term to vanish. As an aside, also the non-mean terms evaluate to zero as $(x_t-\frac12)$ is odd about $\frac12$, $\log(x(1-x))$ is even, and $P$ is symmetric.
\end{enumerate}

Putting this together the FIM of the joint density is given by a sum of the FIM of the stationary density and a coupling term
\begin{align*}
    H_{ij}^{joint} =&~ 2H_{ij}^{stat} + I(\tau)\frac{\partial \log\Delta}{\partial \log \Phi_i}\frac{\partial \log\Delta}{\partial \log \Phi_j}\\
    \approx&~ 2H_{ij}^{stat} + e^{-4\rho\tau}\frac{\partial \log\Delta}{\partial \log \Phi_i}\frac{\partial \log\Delta}{\partial \log \Phi_j}
\end{align*}
As $\tau\to\infty$ this recovers the structure of the FIM based only on the stationary distribution.
Writing $\beta = \partial\log\Delta/\partial\log\Phi$ for the log-parameter gradient of the coupling, as in Eq.~\eqref{eq:joint_hessian}, with components $\beta_\rho$ and $\beta_\mu$, and substituting $2H^{stat} = \lambda_1^{stat} \left[\begin{smallmatrix} 1 & -1 \\ -1 & 1\end{smallmatrix}\right]$, the joint Hessian becomes
\begin{align*}
H_{ij}^{joint} =&~ \lambda_1^{stat}\begin{bmatrix} 1 & -1 \\ -1 & 1 \end{bmatrix} + e^{-4\rho\tau}\beta\beta^\top\\
\end{align*}

\section{Eigendecomposition of the joint density FIM}\label{appx:joint_density_eigdecomp}
The fixed lag Hessian (with the $n=1$ truncation) is a sum of two rank-one matrices.
It can be written as 
\begin{align*}
    H^{joint} = \begin{bmatrix}
        \lambda_1^{stat} + e^{-4\rho\tau}\beta_\rho^2 & -\lambda_1^{stat} + e^{-4\rho\tau}\beta_\rho\beta_\mu\\[1mm]
        -\lambda_1^{stat} + e^{-4\rho\tau}\beta_\rho\beta_\mu & \lambda_1^{stat} + e^{-4\rho\tau}\beta_\mu^2
    \end{bmatrix}.
\end{align*}
such that one can solve for the eigenvalues using the trace and determinant.
The trace is given by 
\begin{align*}
    \mathrm{tr}\,H^{joint} =&~ 2\lambda_1^{stat} + e^{-4\rho\tau}\bigl(\beta_\rho^2 + \beta_\mu^2\bigr).
\end{align*}
and the determinant by
\begin{align*}
    \det H^{joint} =&~ \bigl(\lambda_1^{stat} + e^{-4\rho\tau}\beta_\rho^2\bigr)\bigl(\lambda_1^{stat} + e^{-4\rho\tau}\beta_\mu^2\bigr)\\
                    &- \bigl(-\lambda_1^{stat} + e^{-4\rho\tau}\beta_\rho\beta_\mu\bigr)^2\\
=&~ e^{-4\rho\tau}\lambda_1^{stat}\bigl(\beta_\rho^2 + \beta_\mu^2 + 2\beta_\rho\beta_\mu\bigr)\\
 =&~ e^{-4\rho\tau}\lambda_1^{stat}\bigl(\beta_\rho + \beta_\mu\bigr)^2\\
 =&~ e^{-4\rho\tau}\lambda_1^{stat}\left(\left(\frac{2\rho}{2\rho+\mu} - 2\rho\tau\right) - \frac{2\rho}{2\rho+\mu}\right)^2\\
=&~ \lambda_1^{stat}\,(2\rho\tau)^2\,e^{-4\rho\tau}.
\end{align*}
as the $(\lambda_1^{stat})^2$ terms and the $e^{-8\rho\tau}\beta_\rho^2\beta_\mu^2$ terms both cancel, leaving no higher-order contribution in $e^{-2\rho\tau}$.

The exact eigenvalues are the roots of $\lambda^2 - (\mathrm{tr}\,H^{joint})\lambda + \det H^{joint} = 0$, where the $+$ root is the stiff $\lambda_1^{joint}$ and the $-$ root the sloppy $\lambda_2^{joint}$,
\begin{align*}
    \lambda_{1,2}^{joint} =&~ \tfrac12\left(\mathrm{tr}\,H^{joint} \pm \sqrt{(\mathrm{tr}\,H^{joint})^2 - 4\det H^{joint}}\right),
\end{align*}
with eigenvectors obtained from the first row of $(H^{joint} - \lambda I)v = 0$,
\begin{align*}
    v_{1,2}^{joint} \propto \begin{bmatrix} H^{joint}_{\rho\mu}\\[1mm] \lambda_{1,2}^{joint} - H^{joint}_{\rho\rho}\end{bmatrix}.
\end{align*}

As $\tau\to\infty$ the off-diagonal weight $e^{-4\rho\tau}$ vanishes: the determinant tends to zero while the trace tends to $2\lambda_1^{stat}$.
The stiff root tends to twice the stationary eigenvalue and the sloppy root to zero,
\begin{align*}
    \lambda_1^{joint} \to&~ 2\lambda_1^{stat}, & \lambda_2^{joint} \to&~ 0.
\end{align*}
Since $H^{joint}_{\rho\mu}\to-\lambda_1^{stat}$ and $H^{joint}_{\rho\rho}\to\lambda_1^{stat}$, the eigenvectors approach the stationary stiff and zero-eigenvalue directions,
\begin{align*}
v_1^{joint} \to \begin{bmatrix} -1\\ 1\end{bmatrix}=v_1^{stat}, && v_2^{joint} \to \begin{bmatrix} 1\\ 1\end{bmatrix} = v_2^{stat},
\end{align*}

Thus, for the joint case, the eigenvectors are similar to the stationary case but each rotated from these axes by an angle of order $e^{-4\rho\tau}$. 
The fixed lag thus largely retains the stationary eigenvectors, but with a non-zero sloppy eigenvalue  of $2\rho^2\tau^2 e^{-4\rho\tau}$, driven entirely by the component $\beta_\rho + \beta_\mu = -2\rho\tau$ of the gradient along $(1,1)^\top$.
Note that the $n=1$ truncation result here attains a maximum at $\tau=1/(2\rho)$, whereas the untruncated $\lambda_2^{joint}$ decreases monotonically in $\tau$ and tends to a finite limit as $\tau\to0$.

\section{Bandwidth sensitivity of the fixed-lag eigenvalues}\label{appx:joint_bandwidth}

\begin{figure}[htpb]
    \centering
    \includegraphics[width=\columnwidth]{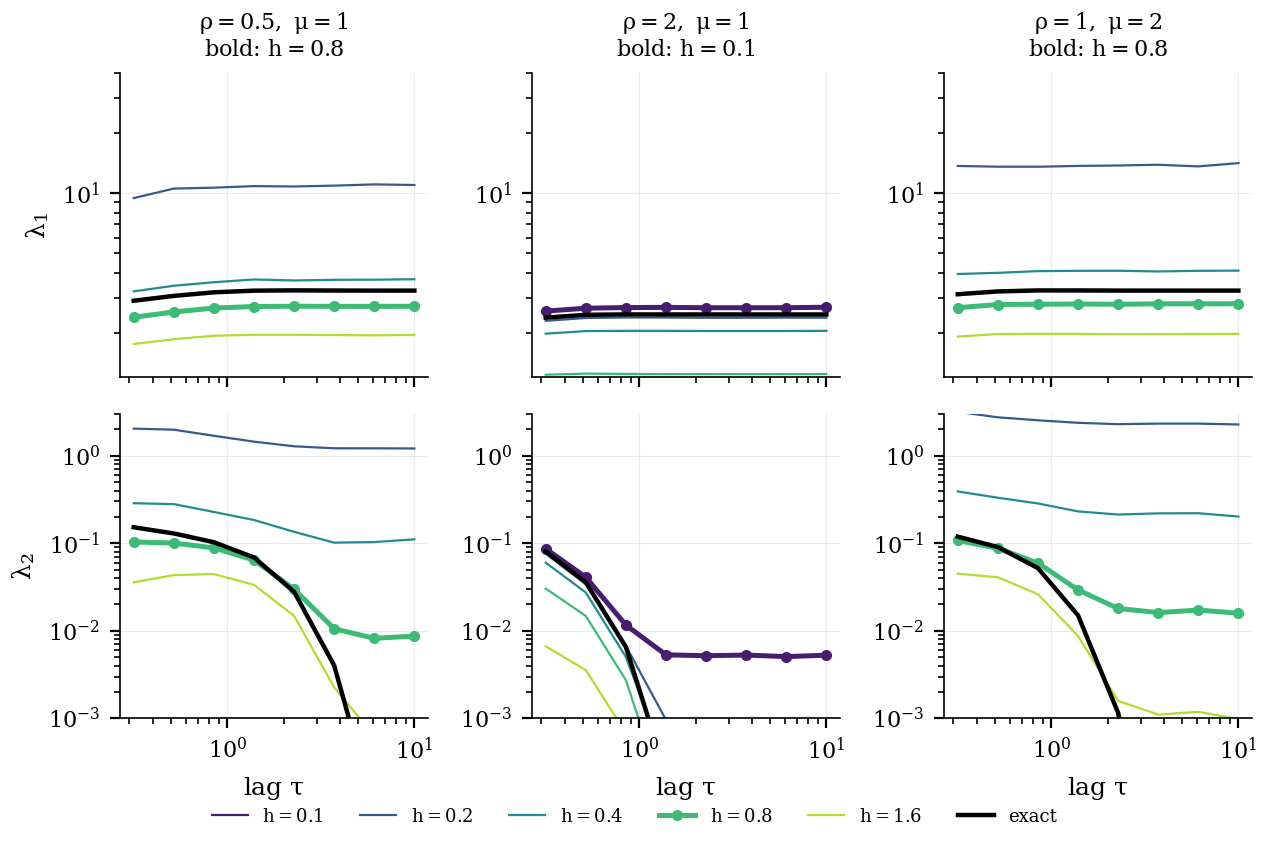}
    \caption{Bandwidth sensitivity of the fixed-lag eigenvalues at the longest sample length, $N = 15000$.
        The layout matches Fig.~\ref{fig:joint_eigenvalues}, with $\lambda_1$ in the top row and $\lambda_2$ in the bottom row, and black the exact spectral result.
        The heavy line marks the bandwidth used in the main text.
        }
    \label{fig:joint_bandwidth}
\end{figure}

\section{Estimated joint density against the slowest-mode form}

\begin{figure}[htpb]
    \centering
    \includegraphics[width=\columnwidth]{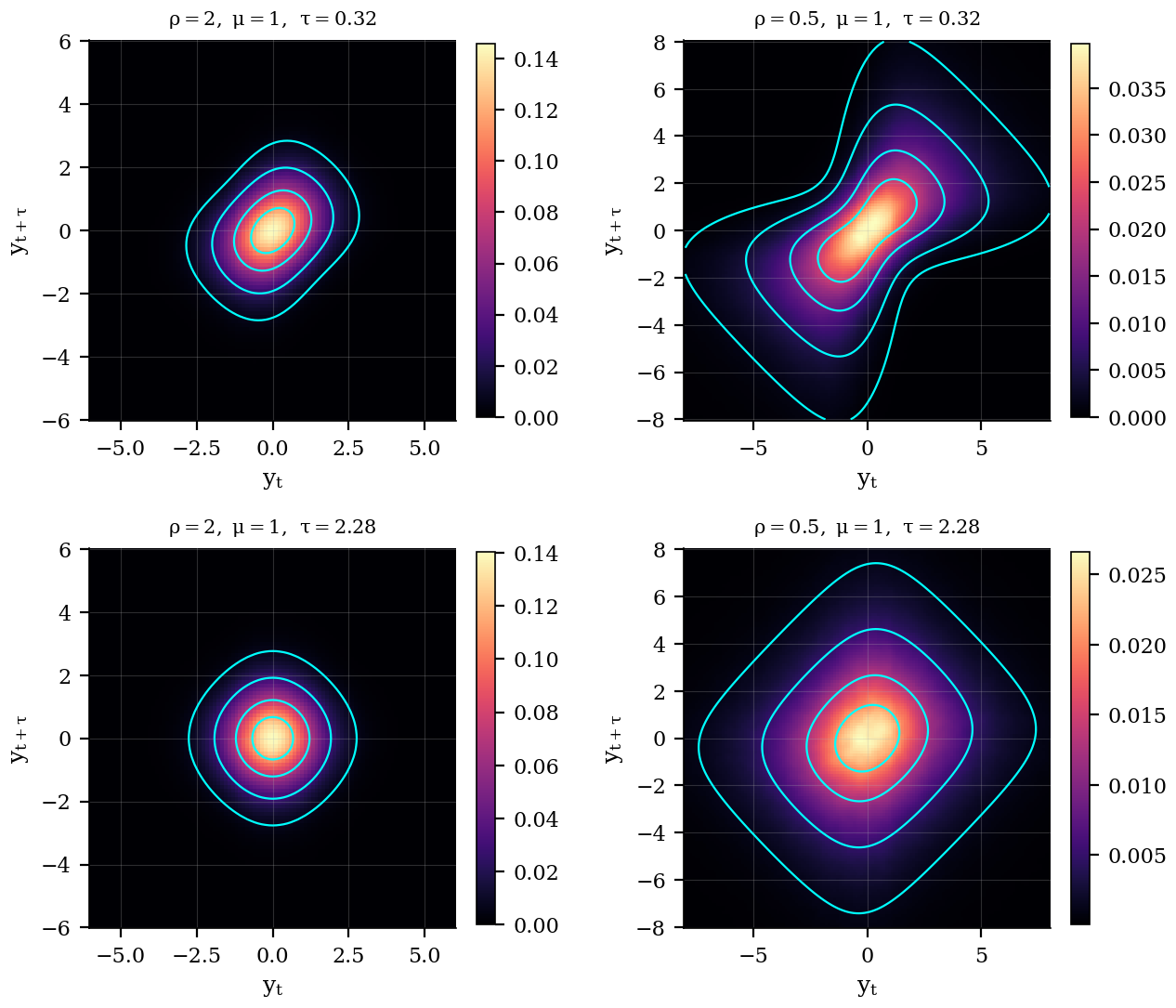}
    \caption{Estimated joint density of the logit pair $(y_t, y_{t+\tau})$ in colour, with contours of the analytical slowest-mode form of Eq.~\eqref{eq:joint_density_n1} in cyan.
        Columns give two parameter points and rows two lags.
        Colour scales differ by panel, since the density of the bimodal point concentrates near the boundaries.}
    \label{fig:joint_density_fig}
\end{figure}

\end{document}